\documentclass[twocolumn,showpacs,preprintnumbers,amsmath,amssymb]{revtex4}
\usepackage[dvips]{graphicx}
\usepackage{color}

\def\ket#1{  \left\vert  #1   \right\rangle   }
\def\bra#1{  \left\langle  #1   \right\vert   }

\begin{document}

\title{Percolation Transition Control in Quantum Networks}

\author{Michael Siomau}

\affiliation{Dubai Men’s Campus, Higher Colleges of Technology, 15825 Dubai, UAE}

\date{\today}

\begin{abstract}
Percolation theory allows simple description of the phase transition based on the scaling properties of the network 
clusters with respect to a single parameter – site/bond occupation probability. How to design a network exhibiting 
the percolation transition for a chosen occupation probability has been an open problem. At the same time, the task 
to find a structurally simple network having the desired property seemed to be impossible. I suggest a model, where
the combination of the classical and the quantum resources creates a percolation transition, continuous or discontinuous 
on demand, for any pre-selected occupation probability and already in the simplest possible one-dimensional network.
\end{abstract}


\maketitle

Quantum networks represent the next generation of complex dynamical structures for communication and advanced
information processing \cite{Kimble}. The quantum features change some aspects of classical network science beyond recognition \cite{Perc, Siomau} leaving many others consistent. An attempt to apply the simplest dynamical theory of percolation \cite{Stauffer} to quantum networks was done more than a decade ago leading to the notion of entanglement percolation \cite{Acin}. This approach was very artificial and didn’t study the scaling behavior of the network clusters near the percolation transition point, thus didn’t advance our understanding of the percolation process in quantum networks. But, a natural and comprehancive extension of the percolation theory to quantum networks is in practical need as quantum networks are becoming reality. In this work, I shall execute standard analysis dictated by the classical percolation theory on a quantum network to see how the \textit{quantumbess} of the network change it's global behaviour. I'll define the fundamental function of any percolation process - the cluster number, and find the critical exponents that govern the percolation near the transition point.

Let me start by considering the direct analog of the classical site percolation in a quantum one-dimensional (1D) network. In the network, the nodes are placed on a line at equal distances and are connected pairwise by quantum channels, those may be open for communication of photons with probability  $0>p\geq 1$ or closed $p=0$. The task is to establish faithful communication between far ends of the network. To accomplish the communication between two chosen nodes having $n-1$ nodes in between, all the channels separating the chosen nodes must be open for communication. The probability to connect the nodes with the open channels is thus $p^n$. The probability to connect the end nodes of the infinite network is  $\lim_{n\rightarrow \infty} p^n = 0$, unless $p=1$.

Apart from the classical strategy above, the quantum networks also allow an alternative way to establish the communication between the far ends by sharing a perfect entanglement pair between them. As it is quite difficults to establish the perfect entanglement over a long distance, it is more practical to assume that initially the entanglement was distributed in the network between any two nodes connected by a channel. To be even more specific, let me assume the a pair of entangled qubits initially prepared into the  two-qubit state $\ket{\phi}=\sqrt{\tau}\ket{00}+\sqrt{1-\tau}\ket{11}$ was distributed between each pair of nodes connected by a channel. Here the states are represented in the computational basis. If $\tau=1/2$ the state is called maximally entangled or perfect. There are few strategies to communicate in a quantum network with the entangled states over long distances. A strategy based on the entanglement swapping was considered in \cite{Acin} concluding that the communication between the far ends cannot be established unless the state $\ket{\phi}$ is maximally entangled. Another strategy may be to convert the imperfect state $\ket{\phi(\tau \neq 1/2)}$ into the maximally entangled by applying the filtering operation $F=\sqrt{\tau}\ket{0}\bra{0} + \sqrt{\tau}\ket{1}\bra{1}$ \cite{Tiersh}. The filtering results in the perfect entanglement pair with probability  $p_e=2\tau(1-\tau)$, which never exceeds $1/2$. Hence, the probability to connect the end nodes of the infinite network is $\lim_{n\rightarrow \infty} p_e^n = 0$. Conclusively, if in the quantum 1D network the imperfect entanglement pairs are distributed between the nodes, the probability to establish the communication between the far ends of the network is zero.

Suppose now that we have a quantum 1D network with some channes open for communication and some amount of entaglement equally distributed between the neighboring nodes. Will the combination of such classical and quantum resources change the percolation properties of the network? To answer this question, let me assume that at the initial moment of time each pair of neighboring nodes shares a pair of qubits prepared into the two-qubit state $\ket{\phi(\tau\neq1/2)}$, which may be converted into the maximally entangled state via filtering with probability $p_e$ and all the channels are closed for communication, i.e. $p=0$. Next, let $p>0$ channels of the network to be open for communication and the entangled pairs are filtered. The processes of the channel opening and the filtering of the entangled pairs are statistically independent, hence both open channels and perfect entanglement pairs contribute independently to our ability to communicate information. The \textit{communication cluster} in the quantum network may be defined as a combination of open channels and perfect entanglement pairs converted over closed channels. The perfect entanglement pairs over the open channels are ignored, as they add nothing to our ability to propagate information. A communication $s$-cluster may be defined as an agglomeration connecting $s+1$ nodes by a combination of $i$ open channels and $s-i$ perfect entanglement pairs over closed channels. The cluster number is $n_s=p^i p_e^{s-i}$. There is a significant variation in the combinations of the open channels and the perfect entangled pairs in a communication cluster, which may be accounted by the binomial coefficient $\binom{s}{i}$. Thus, the probability that an arbitrary open channel or a perfect entanglement pair is part of a communication $s$-cluster is 

\begin{equation}
sn_s=\sum_{i=1}^s \binom{s}{i} p^i p_e^{s-i} (1-p)^2(1-p_e)^2 \, .
\nonumber
\end{equation}

Directly computed mean cluster size $S=\sum_{s=1}^\infty  \frac{s}{\sum_{s=1}^\infty sn_s} sn_s$ diverges at $p_c=1-p_e$ indicating the percolation transition. Indeed, the perfect entanglement pairs cover $p_e$ fraction of the closed network channels, while $p$ channels are open for communication. If $p_e+p \geq 1$, there is a nonzero probability to establish the communication between the end nodes. In other words, the communication cluster is macroscopic with respect to the network size.

Explicit definition of $sn_s$ allows analytial computation of all the critical exponents governing the behaviour of the quantum network before the percolation transition point. Near the the percolation transition poin $p_c$ the mean cluster size scales as $S \sim (p_e - p)^{-1}$  with the critical exponent $\gamma=1$. The characteristic cluster size scales as $(p_e-p)^{-1}$, thus the critical exponent $\sigma=1$. The pair connectivity $g(r)=(p+p_e)^r$ gives us the  exponent $\nu=1$. The Fisher exponent is $\tau=2$. 

\begin{figure}
\centering
\includegraphics[width=.8\linewidth]{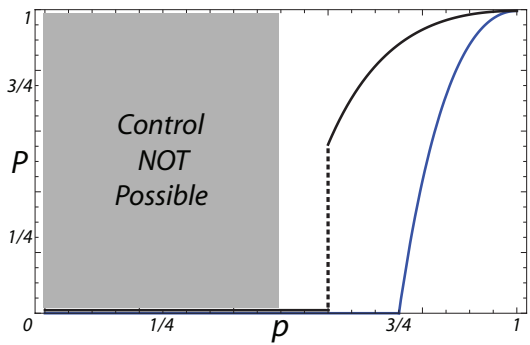}
\caption{Continious percolation transition for $p=0.75$ and $p_e=0.25$, and discontinious percolation transition with the delayed filtering for $p=0.6$ and $p_e=0.49$. }
\label{fig}
\end{figure}

The percolation strength or the probability to connect the end nodes of the infinite network, which describes the behavior of the giant component after the percolation transition, may be obtained from the following considerations. Let me define the percolation strength as $P=1-Q^2 Q_e^2$, where  $Q=(1-p)+pQQ_e$ is the probability that a chosen open channel is not connected to the infinity by an open channel nor by the perfect entangled pair and  $Q_e=(1-p_e )+p_e QQ_e$ is the probability that a chosen perfect entangled pair is not connected to the infinity by an open channel nor by the perfect entangled pair. Solving these equations together we have 

\begin{equation}
P=1-\frac{(1-p)^2(1-p_e)^2}{p^2 p_e^2} \, .
\nonumber
\end{equation}
Any $p_e\neq0$ corresponds to a percolation transition at $p_c=1-p_e$. Taking into account that  $p_e$ is restricted by the filtering operation as $p_e<1/2$, the percolation transition can be called on demand for any $p_c>1/2$ as exemplified in Fig.\ref{fig}. The equation for the percolation strength $P$ scales after the percolation transition as $P\sim (p-p_c)^1$ giving the critical exponent  $\beta=1$. All together the critical exponents give us the full description of the cluster scaling in the quantum network.

In the classical percolation theory, the critical exponents are connected to each other by the fundamental relations called the scaling laws. For example, $\beta=(\tau-2)/\sigma$. The communication clusters in the quantum network exhibit the same scaling before the percolation transition as the clusters in the classical 1D percolation. But, the fact that there is a non-trivial percolation transition violates the fundamental scaling laws valid for all classical networks.

If we allow the processes of channel opening and the filtering to progress in parallel, the percolation transition is continuous. But, the filtering operation may be delayed and released at any time afterwards. For example, let $p_e=0.49$ and the filtering is delayed, than for any $p<1$ the network doesn’t percolate. If for $p=0.6$ we allow the filtering, there is an instantaneous jump in the percolation strength from $P=0$ to $P\approx0.55$. This jump has pure operational meaning and should not be confused with the explosive percolation \cite{DSouza}, which is the concequence of abnormal scaling.

The analysis of the scaling properties of the quantum 1D network may be extended to a quantum network with arbitrary structure, which combines the classical and the quantum resourses. If the latter resource is interpreted as the initial condition for percolation process, would it be correct to say that the statistical future of a quantum network is defined by it’s quantum past? In any case, the percolation transition control seems to be impossible task if we have only classical or quantum resources, but can be easily handled by their combination. This fact may substantially advance our understanding of the complex quantum structures and radically change the approaches for quantum information processing.

\end{document}